\documentclass[nofootinbib,
 reprint,
 amsmath,amssymb,
 aps,prd,floatfix]{revtex4-1}

\usepackage{graphicx}
\usepackage{dcolumn}
\usepackage{bm}
\usepackage{amsthm,amsbsy,amsfonts,amscd}
\usepackage{bm}
\usepackage{color}
\definecolor{dark_green}{RGB}{26,137,34}
\usepackage[utf8]{inputenc} 
\begin{document}
\interfootnotelinepenalty=10000
	\title{Cosmology in the laboratory: An analogy between hyperbolic metamaterials and the Milne universe}
	\author{David Figueiredo}
    \affiliation{Departamento de F\'isica, CCEN, Universidade Federal da Para\'iba,
Caixa Postal 5008, 58051-900, Jo\~ao Pessoa, Para\'iba, Brazil }
    
    \author{Fernando Moraes}
\altaffiliation[Also at ]{Departamento de F\'isica, Universidade Federal da Para\'iba,
Caixa Postal 5008, 58051-900, Jo\~ao Pessoa, Para\'iba, Brazil }
\affiliation{%
 Departamento de F\'isica, Universidade Federal Rural de Pernambuco,\\
52171-900, Recife, Pernambuco, Brazil
}
\author{S\'ebastien Fumeron and Bertrand Berche}
\affiliation{ Statistical Physics Group, Laboratoire de Physique et Chimie Th\'eoriques, Universit\'e de Lorraine,
    B.P. 70239,\\  54506 Vand\oe uvre les Nancy, France}
    
\title{Cosmology in the laboratory: an analogy between hyperbolic metamaterials and the Milne universe}{}

\date{\today}

\begin{abstract}
This article shows that the compactified Milne universe geometry, a toy model for  the big crunch/big bang transition, can be realized in hyperbolic metamaterials, a new class of nanoengineered systems which have recently found its way as an experimental playground for cosmological ideas. On one side, Klein-Gordon particles, as well as tachyons, are used as probes of the Milne geometry. On the other side, the propagation of light in two versions of a liquid crystal-based metamaterial provides the analogy. It is shown that ray and wave optics in the metamaterial mimic, respectively, the   classical trajectories and wave function propagation, of the Milne probes, leading to the exciting perspective of realizing experimental tests of particle tunneling through the cosmic singularity, for instance.

\end{abstract}

\maketitle
\section{Introduction}
Initial conditions are always a trouble in cosmology but can be circumvented by cyclic universe models, like an endless repetition of big crunches followed by big bangs, for instance. This is in fact an old idea that can be traced back to ancient mythologies. Even without referring to initial conditions some issues remain in these models, notably the passage through the singularity, the transition from big crunch to big bang. Recently, safe transition has been proposed \cite{from_BC_to_BB,steinhardt2002cosmic}, where the singularity is nothing more than the temporary collapse of a fifth dimension. The three space dimensions remain large and time keeps flowing smoothly. A toy model for the geometry of this transition is the compactified 2D Milne universe \cite{horowitz1991singular} which is essentially a double cone in 3D Minkowski spacetime. Because there is still a singularity in one spatial dimension, a physically correct model should be able to describe the propagation of a particle through it. An example of such a model \cite{poloneses2006} revealed that the compactified Milne universe seems to model the cosmic singularity in a satisfactory way. It is important to stress that the collapse here is far less severe than in ordinary 4D general relativity, because it happens just in one spatial dimension, whereas in the ordinary case the collapse happens in the entire four-dimensional spacetime \cite{steinhardt2002cosmic}. The question we want to address in this work is: can we simulate the passage through the singularity in the laboratory? 

Recent advances in the field of metamaterials suggest this possibility. Arising from a pioneering idea by Veselago \cite{Veselago1964} and developed by Pendry \cite{Pendry2006}, metamaterials are artificial media structured at subwavelength scales, such that their permittivity and permeability values can be taylored quasiarbitrarily (for instance, they may exhibit negative refractive index). Because analogy is a powerful tool that a physicist possesses to arrive at an understanding of the properties of nature, along with the fact that visualizations of celestial object features in the laboratory have been a charming subject for human beings over centuries, analogue gravity became an active field in physics with the help of metamaterials. Therefore, several works based on different results of general relativity were done, including topics like analog spacetimes \cite{Dielectric_Spacetime_PRD,Metamaterials_Curved_Spacetime}, time travel \cite{Time_Travel_PRD}, cosmic strings \cite{Mackay_cosmic}, celestial mechanics \cite{Mimicking_Celestial_Mechanics}, black holes \cite{Isabel_Black_Hole,TOptics_mimics_BH} and wormholes \cite{Wormhole_PRL}, just to cite a few. However, most of the approaches aforementioned rely on transformation optics \cite{leonhardt_geolight}, where the permittivity $\epsilon_{ij}$ and permeability $\mu_{ij}$ tensors must be numerically equal (impedance matched). It turns out that sometimes it is difficult to obtain such a constraint experimentally. Nevertheless, a particularly promising class of such media is that of hyperbolic metamaterials, for which one of the eigenvalues of either the permittivity or the permeability tensors do not share the same sign with the two others (thus, they do not need to be impedance matched). Hyperbolic metamaterials are now extensively studied, both for practical purpose (enhanced spontaneous emission, hyperlenses \cite{Poddubny2013,semiclassical_hyperlens,hyperlens2006}) but also for modeling cosmological phenomena (metric signature transition \cite{Smolyaninov2010}, modeling of time \cite{smolyaninov2011modeling,figueiredo2016modeling}, and even inflation \cite{smolyaninov2012experimental}). 

Therefore, since inflationary models of the universe can be mimicked through hyperbolic metamaterials, it can be interesting to see if the cyclic/ekpyrotic ones also can. As far as light propagation is concerned, a ($2+1$) Minkowski spacetime can be simulated with such materials. As will be seen below, the quantum dynamics of Klein-Gordon particles through the big crunch/big bang transition may be experimentally verified with light propagating in a suitable hyperbolic metamaterial.


\section{The $\bm{\mathcal{M}_C}$ universe}
\label{CM_section}
\begin{figure}
	\centering
	\includegraphics[width=.6\columnwidth]{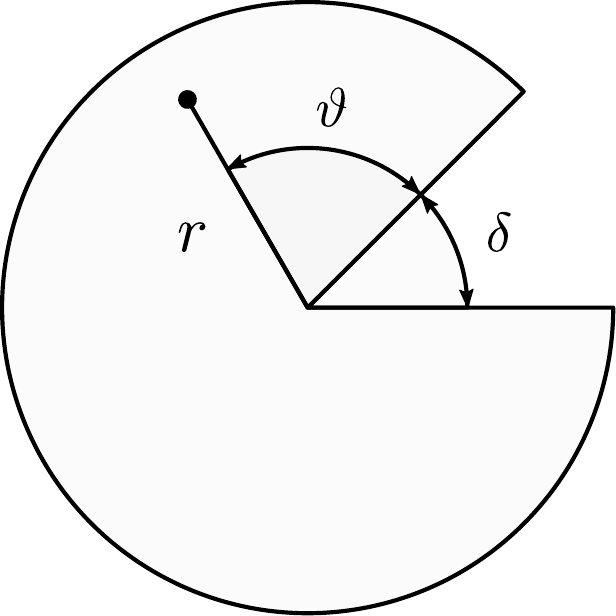}
    \caption{Plane sector which can be bent around a cone. The deficit angle $\delta=2\pi(1-\sin\theta)$ reduces the range of the polar angle $\vartheta$ to $[0,2\pi\sin\theta)$.}
    \label{sector_fig}
\end{figure}

    In this section we will summarize the definition and features of the Milne universe and the compactified Milne universe, $\mathcal{M}_C$. However, first it is useful to analyze the geometry of a rectangular cone in $\mathbb{R}^3$ to get some insight. Therefore, let $\theta$ be the angle between the generatrix and the axis. In Cartesian coordinates $(x,y,z)$ the cone surface will have the following form
\begin{equation}
	x^2+y^2=z^2\tan^2\theta,
	\label{cart_3d_cone}
\end{equation}
with parametric equations
\begin{align}
	x&=r\sin\theta\cos\left(\vartheta/\sin\theta\right), \label{x_mk1}\\
    y&=r\sin\theta\sin\left(\vartheta/\sin\theta\right), \label{y_mk1}\\
    z&=r\cos\theta, \label{z_mk1}
\end{align}
and position vector 
\begin{equation}
	\mathbf{r}=\left[r\sin\theta\cos\left(\vartheta/\sin\theta\right),r\sin\theta\sin\left(\vartheta/\sin\theta\right),r\cos\theta\right],
    \label{position_vector}
\end{equation}
due to the fact that it is locally isometric to a piece of the plane \cite{manfredo1976carmo}. In the parametrization above $0\leq\vartheta<2\pi\sin\theta$ is the polar angle and $2\pi(1-\sin\theta)$ is the deficit angle (see Fig. \ref{sector_fig}). Rescaling the polar angle as $\phi=\vartheta/\sin\theta$ one recovers the parametrization in terms of spherical coordinates $(r,\theta,\phi)$. Thus, the line element at the cone surface is, in terms of $\phi$,
\begin{equation}
	ds^2=dr^2+r^2\sin^2\theta d\phi^2.
    \label{cone_3d_metric}
\end{equation}
Then, because $\theta=\text{constant}$ we have an induced 2D metric on the cone with the following metric tensor
\begin{equation}
	g_{ij}= \begin{pmatrix}
		1 & 0  \\
		0  & r^2\sin^2\theta
	\end{pmatrix},
    \label{cone_metric_tensor}
\end{equation}
where $g_{ij}=\mathbf{e}_{i}\cdot\mathbf{e}_j$ $(\mathbf{e}_i=\partial\mathbf{r}/\partial x^i,x^1=r,x^2=\phi)$.

To extend the above case for the double cone, we modify the radial coordinate $r$ making $r\rightarrow l$, so that $l \in \mathbb{R}$. Thus, $l>0$ for the upper cone and $l<0$ for the lower one. Making these changes in Eq. (\ref{cone_3d_metric}), we obtain the 2D induced line element for the double cone
\begin{equation}
	ds^2=dl^2+l^2\sin^2\theta d\phi^2.\label{doublecone_3d_metric}
\end{equation}

Concerning the cosmological model, consider the usual Robertson-Walker line element with negative spatial curvature \cite{misner1973gravitation}
\begin{equation}
	ds^2=-dt^2+a(t)^2\left(d\chi^2+\sinh^2\chi \,d\Omega^2\right),
	\label{RW_metric}
\end{equation}
where $a(t)$ is the expansion factor of the universe and $d\Omega^2=d\theta^2+\sin^2\theta d\phi^2$ the solid angle. For a linear expansion factor $a(t)=t$, we obtain the Milne universe metric
\begin{equation}
	ds^2=-dt^2+t^2\left(d\chi^2+\sinh^2\chi\,d\Omega^2\right),
	\label{RW_milne}
\end{equation}
which was proposed by Edward Arthur Milne in $1933$ and represents a homogeneous, isotropic and expanding model for the universe \cite{gron2007einstein} which grows faster that simple cold matter dominated or radiation dominated universes. We are interested in the hypersurfaces where $d\Omega=0$. Then, Eq. (\ref{RW_milne}) reduces to
\begin{equation}
	ds^2=-dt^2+t^2d\chi^2.
    \label{M_metric}
\end{equation}

Next, we  make a coordinate transformation to new variables $(T,X)$ given by
\begin{align}
T&=t\cosh\chi,\label{tchange}\\
X&=t\sinh\chi,\label{xchange}
\end{align}
which leads to the following form for the line element in Eq. (\ref{M_metric}), namely
\begin{equation}
	ds^2=-dT^2+dX^2,\label{minkowski_metric}
\end{equation}
which is the usual Minkowskian metric in two dimensions. However, one must stress that the Milne universe only covers half of the Minkowski spacetime. To see why, consider the lines of constant values $\chi=\chi_0$ in Eqs. (\ref{tchange}) and (\ref{xchange}),
\begin{equation}
	X=T \tanh\chi_0.
    \label{chi_const}
\end{equation}	
As a result, we have straight lines in a Minkowskian diagram. Taking the limits $\chi_0\rightarrow\pm\infty$ in the equation above, one gets the light rays $X=\pm T$, which form the boundaries of the past and future light cones from the origin $(0,0)$. Thus, one is confined in this region where $X^2-T^2<0$, since the slope of the line given by Eq. (\ref{chi_const}) goes from $0$ (for $\chi_0=0$) to $\pm 1$ (for $\chi_0\rightarrow\pm\infty$).

Another important point to stress is that the three-dimensional space for the comoving Milne observers has infinite extension. The reason is due to the fact that the lines of constant $t=t_0$ are hyperbolas, namely
\begin{equation}
	T^2-X^2=t_0^2,
    \label{milne_tconst}
\end{equation}
and each one of the hyperbolas has infinite length. This fact is expected since the Milne universe has a negative spatial curvature \cite{gron2007einstein}.

\begin{figure}
	\centering
	\includegraphics[width=1.0\columnwidth]{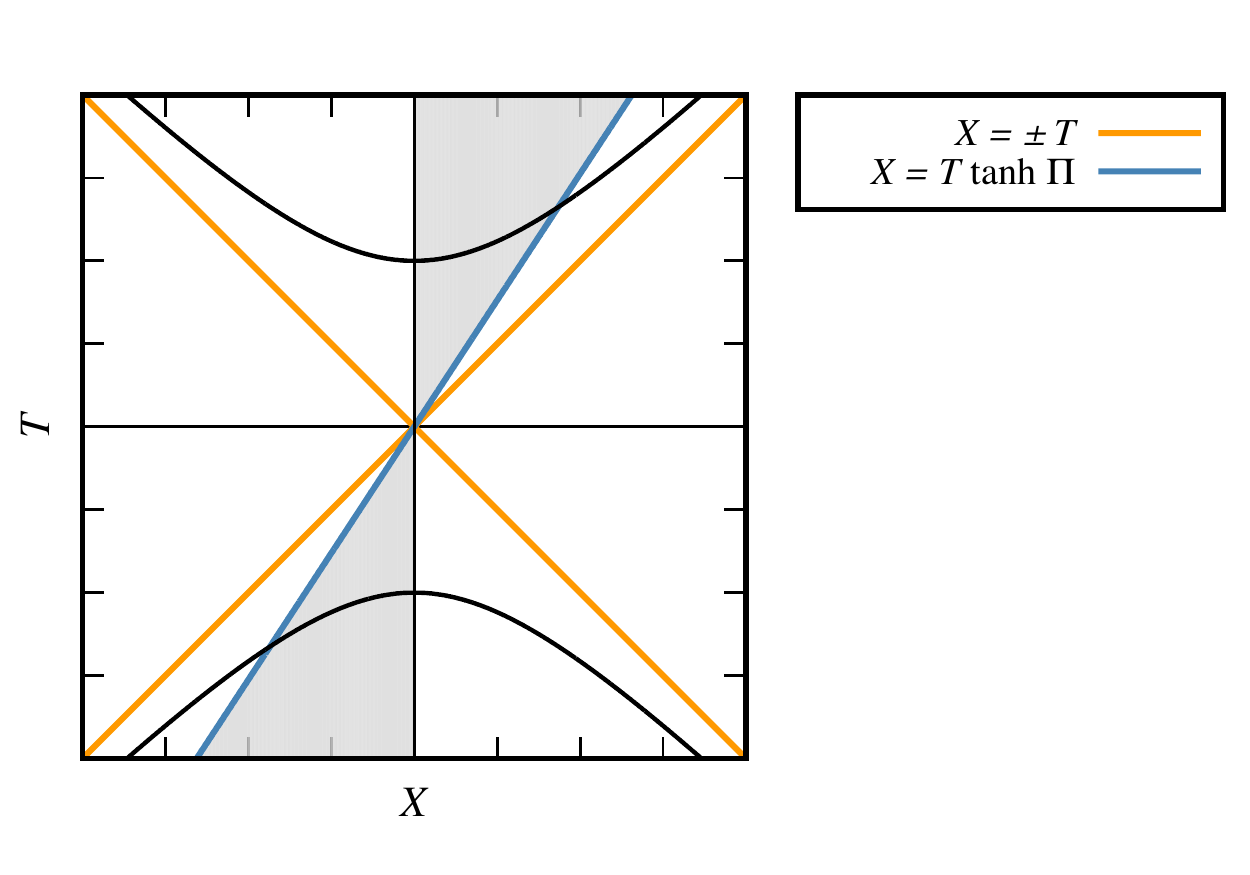}
	\caption{Minkowski diagram showing the $\mathcal{M}_C$ universe (gray region) delimited by the identification between the $T$-axis and the blue line $X=T\tanh \Pi$. The orange lines are showing the light rays $X=\pm T$ while the hyperbolas are surfaces for constant $t$ in Eq.(\ref{milne_tconst}).}
	\label{MC_universefig}
\end{figure}

Therefore, in order to compactify the Milne universe, we follow the usual approach \cite{from_BC_to_BB,steinhardt2002cosmic,horowitz1991singular,tolley2004cosmological} and let the variable $\chi$ acquire some period $\Pi$. The meaning of this is as follows, in the Minkowski diagram $(T,X)$ the lines $X=0$ and $X=T\tanh \Pi$ should be identified as one, for instance. Therefore, because one is constrained between these two lines, the Milne universe now has a finite length and is called compactified Milne universe, $\mathcal{M}_C$ (see Fig. \ref{MC_universefig}).

\begin{figure}
\centering
\includegraphics[width=1.0\columnwidth]{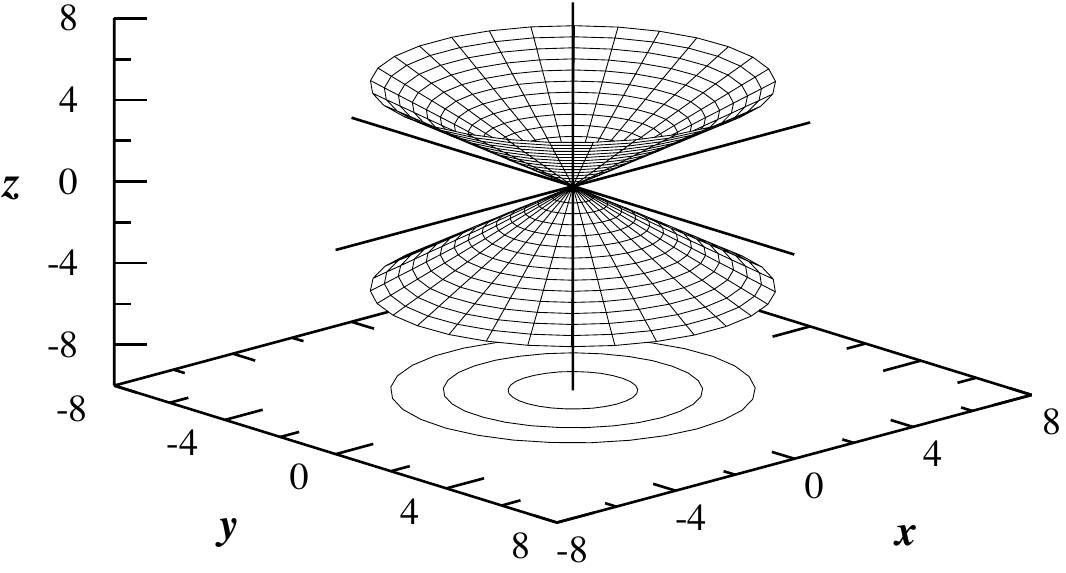}
\caption{Double cone surface corresponding to Eqs. (\ref{cart_3d_cone}) and (\ref{CM_cone}) for 3D Euclidean space and 3D Minkowski space, respectively. The upper half part corresponds to $l,t>0$ while the lower one is for values of $l,t<0$.}
\label{doubleConefig}
\end{figure}

Following Refs. \cite{malkiewicz2006simple,poloneses2006}, one can visualize the $\mathcal{M}_C$ universe through a mapping into a three-dimensional Minkowski space with $ds^2=-dz^2+dx^2+dy^2$, with
\begin{align}
	x&=t\kappa\cos\left(\chi/\kappa\right), \label{x_mk}\\
    y&=t\kappa\sin\left(\chi/\kappa\right), \label{y_mk}\\
    z&=t\sqrt{1+\kappa^2}, \label{z_mk}
\end{align}
where $t \in \mathbb{R}^1$ and $0<\kappa \in \mathbb{R}^1$ is a constant parameter for compactifications (in Refs. \cite{malkiewicz2006simple,string_theory_book} it is shown that $\kappa$ is related to the rapidity $\tanh^{-1}v=2\pi \kappa$ of a finite Lorentz boost). Thus, without loss of generality, we take the period of $\chi$ as $2\pi \kappa$. Moreover, as in the previous case of the ordinary cone, we rescale $\chi$ by $\phi=\chi/\kappa$ (and hence giving a period of $2\pi$ for $\phi$). Solving Eqs. (\ref{x_mk})--(\ref{z_mk}) one gets the following constraint equation
\begin{equation}
	x^2+y^2=\left(\frac{\kappa^2}{1+\kappa^2}\right)z^2.
	\label{CM_cone}
\end{equation}
This equation is similar to Eq. (\ref{cart_3d_cone}) because the space is Euclidean for the planes $z=\text{constant}$ and also due to the periodicity of $\phi$. As $x,y,z$ can assume both positive and negative values, Eq. (\ref{CM_cone}) represents a double cone with a vertex at $(0,0,0)$ in the 3D Minkowski space (see Fig. \ref{doubleConefig}). However, due to the timelike aspect of $z$, the cone angle $\theta$ is a hyperbolic angle with $\tanh^2\theta=\kappa^2/1+\kappa^2$.

Taking into account  all previous parametrizations, the parametric equations (\ref{x_mk})--(\ref{z_mk}) become
\begin{align}
	x&=t\sinh\theta\cos\phi, \label{x_mk2}\\
    y&=t\sinh\theta\sin\phi, \label{y_mk2}\\
    z&=t\cosh\theta, \label{z_mk2}
\end{align}
where $\sinh\theta=\kappa$, $\cosh\theta=\sqrt{1+\kappa^2}$ and $\phi\in\mathbb{S}^1$. As for the metric in $\mathcal{M}_C$,
\begin{equation}
	ds^2=-dt^2+\kappa^2t^2d\phi^2.
    \label{milne_metric}
\end{equation}

The presence of $\kappa^2$ in the above metric indicates a conic singularity of the curvature  at the origin. As we will see below, this has important implications to the geodesics and wave functions of particles approaching the singularity, which acts as a filter  for classical particles and a phase eraser for quantum ones. Due to the fact that the ordinary cone surface is embedded in three-dimensional Euclidean space and the $\mathcal{M}_C$ universe has a cone surface embedded in three-dimensional Minkowski space, they share some similarities. For instance, the Milne metric tensor
\begin{equation}
	g_{ij}= \begin{pmatrix}
		-1 & 0  \\
		0  & \kappa^2t^2
	\end{pmatrix}.
\end{equation}
is similar to the two-dimensional induced metric tensor given by Eq. (\ref{cone_metric_tensor}). Also, as a last comparison, let us consider the Laplace-Beltrami operator, which will be used later in this paper. Thus,
\begin{equation}
	\Delta=\nabla_i\nabla^i=\frac{1}{\sqrt{\left|g\right|}}\partial_i\left(\sqrt{\left|g\right|}g^{ij}\partial_j\right).
    \label{laplace_beltrami}
\end{equation}
where $g^{ij}$ is the inverse metric tensor and $g=\det(g_{ij})$. From Eqs. (\ref{doublecone_3d_metric}) and (\ref{milne_metric}), we will have
\begin{align}
	\Delta_{\text{Cone}}&=\frac{1}{l}\frac{\partial}{\partial l}\left(l\frac{\partial}{\partial l}\right)+\frac{1}{l^2\text{sin}^2\theta}\frac{\partial^2}{\partial\phi^2},\label{laplacian_Cone}\\
	\Delta_{\mathcal{M}_C}&=-\frac{1}{t}\frac{\partial}{\partial t}\left(t\frac{\partial}{\partial t}\right)+\frac{1}{\kappa^2t^2}\frac{\partial^2}{\partial\phi^2}, \label{laplacian_CM}
\end{align}
which also look similar due to the resemblances between the two metrics, aside from the timelike behavior in $t$ and the hyperbolic cone angle factor $\kappa^2=\sinh^2\theta$.

\section{Classical particle in $\bm{\mathcal{M}_C}$}
\label{classical_MC}
In this section we will obtain the path followed by free classical particles (timelike geodesics) in  $\mathcal{M}_C$. Thus, we start with the relativistic action (in natural units $c=1$) given by \cite{landau2} 
\begin{equation}
	 S=-m\int\sqrt{-g_{ij}\dot{x}^{i}\dot{x}^{j}}d\lambda, \label{action1}
\end{equation}
where the dot notation stands for a derivative with respect to an affine parameter $\lambda$ along the curve (which can be the proper time) and $\mathcal{L}(x^i,\dot{x}^i,\lambda)=-m\sqrt{-g_{ij}\dot{x}^{i}\dot{x}^{j}}$ is the Lagrangian. The variation of the action, $\delta S$, will be
\begin{equation}
	\delta S=-m\,\delta \int\sqrt{-g_{ij}\dot{x}^{i}\dot{x}^{j}}d\lambda, \label{actionvar}
\end{equation}
thus,  the least action principle $\delta S=0$ demands that
\begin{equation}
	\delta \int\sqrt{-g_{ij}\dot{x}^{i}\dot{x}^{j}}d\lambda=0, \label{actionvar2}
\end{equation}
from which the geodesic equations follow
\begin{equation}
	\ddot{x}^i+\Gamma^i_{jk}\dot{x}^j\dot{x}^k=0,\label{geodesic2}
\end{equation} 
where $\Gamma^i_{jk}$ are the Christoffel symbols of the second kind.

An alternative, and more direct route to get the geodesic equations and Christoffel coefficients is through the Euler-Lagrange equations for the kinetic Lagrangian
\begin{equation}
	\mathcal{L}_\text{kin}=\frac{1}{2}\,g_{ij}\dot{x}^i\dot{x}^j,
    \label{K_lagrange}
\end{equation}
which for the Milne metric (\ref{milne_metric}) gives the following equations
\begin{align}
	\ddot{\phi}+\frac{2}{t}\,\dot{t}\,\dot{\phi}&=0, \label{geo_eq_p}\\
	\ddot{t}+\kappa^2t\,\dot{\phi}^2&=0.\label{geo_eq_t}
\end{align}
From the equations above, one can see that the nonvanishing Christoffel symbols are $\Gamma^{\phi}_{t\phi}=\Gamma^{\phi}_{\phi t}=1/t$ and $\Gamma^{t}_{\phi\phi}=\kappa^2t$. Furthermore, since $\phi$ is a cyclic coordinate in the Lagrangian, $\partial \mathcal{L}_\text{kin}/\partial \phi=0$ and the angular momentum $p_{\phi}=\partial \mathcal{L}_\text{kin}/\partial \dot{\phi}=\kappa^2t^2\dot{\phi}$ is conserved. This fact is expressed in Eq. (\ref{geo_eq_p}).

Substituting $p_{\phi}=\kappa^2J=\text{constant}$ in Eq. (\ref{geo_eq_t}) and after some manipulations, the second geodesic equation becomes
\begin{equation}
\dot{t}^2-\frac{\kappa^2J^2}{t^2}=2E,
\label{geot_almost}
\end{equation}
where $E$ is a constant of integration. The left-hand side of Eq. (\ref{geot_almost}) is nothing else but minus the square modulus of the velocity $u^i=\dot{x}^i$. As a result,  recalling that in special relativity one has $u^iu_i=-1$ and $\lambda$ is an affine parameter, it is useful to choose $u^iu_i=-2E=-1$. Then, Eq. (\ref{geot_almost}) becomes
\begin{equation}
\dot{t}^2-\frac{\kappa^2J^2}{t^2}=1.
\label{geot_almost2}
\end{equation}
Solving Eq. (\ref{geot_almost2}) for $\dot{t}$, we get
\begin{equation}
\dot{t}=\pm \sqrt{\frac{t^2+\kappa^2J^2}{t^2}},
\end{equation}
which leads to a simple integration of the form
\begin{equation}
\int{\frac{tdt}{\sqrt{t^2+\kappa^2J^2}}}=\pm\int{d\lambda},
\end{equation}
and therefore to the parametric equation $t=t(\lambda)$ given by
\begin{equation}
	t(\lambda)=\pm\sqrt{\Delta\lambda^2-\kappa^2J^2},
    \label{parametric_t}
\end{equation}
where $\Delta\lambda=\lambda -\lambda_0$, with $\lambda_0$ being a constant of integration.

\begin{figure}
\centering
\includegraphics[width=1.0\columnwidth]{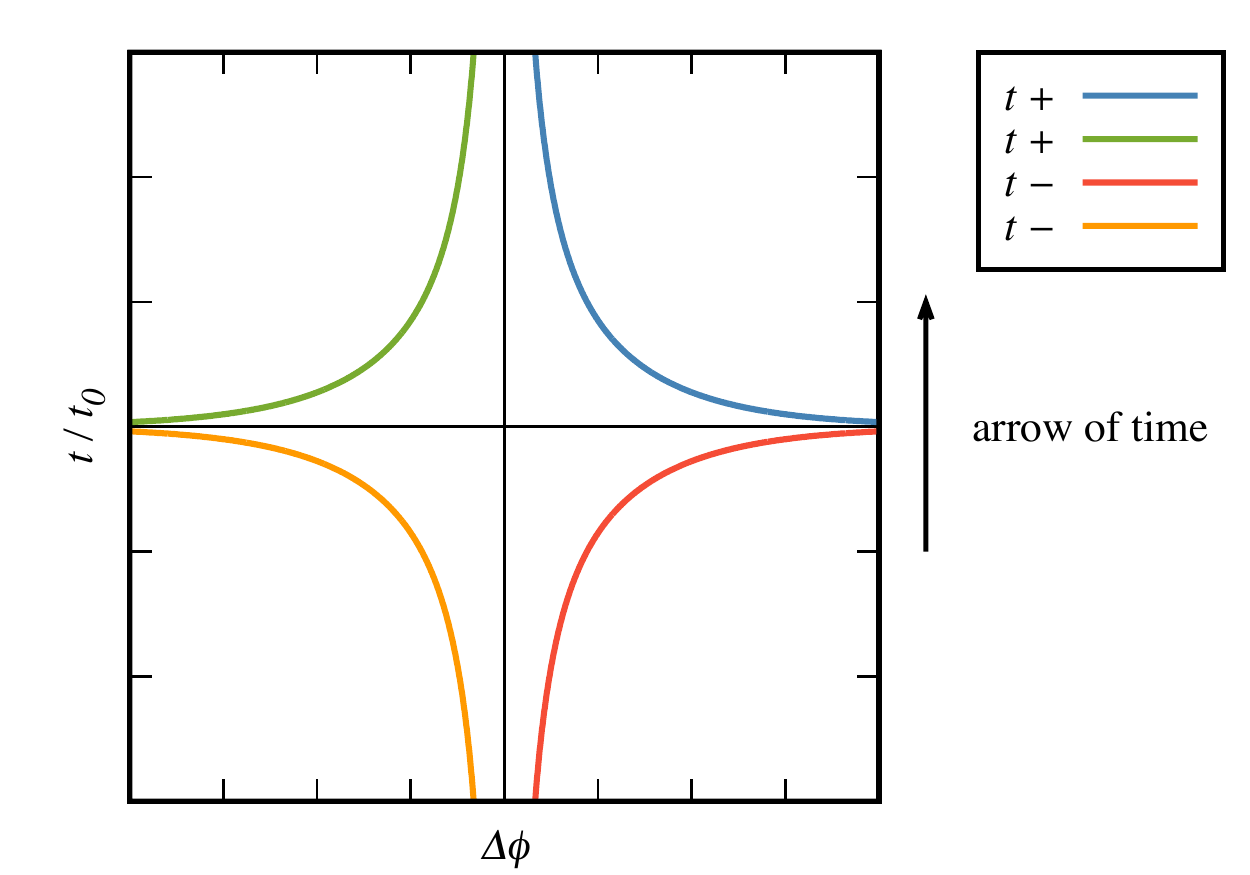}
\caption{Graph for the geodesic equation given by Eq. (\ref{trajec_eq}) with $t$ in units of $t_0$. The blue and green (orange and red) lines correspond to trajectories in the upper (lower) cone. The arrow of time pointing up shows that time always increases, which means that in the first and third quadrants the angle $\phi$ decreases with time ($J<0$), while in the second and fourth ones it increases with time ($J>0$).}
\label{time_geodesic_fig}
\end{figure}

In order to obtain the parametric equation  $\phi=\phi(\lambda)$, we substitute Eq. (\ref{parametric_t}) in the relation $t^2\dot{\phi}=J$ and solve for $\dot{\phi}$. Namely,
\begin{equation}
	\dot{\phi}=\frac{J}{\Delta\lambda^2-\kappa^2J^2},
\end{equation}
leading to the following integration
\begin{equation}
	\int{d\phi}=J\int{\frac{d\lambda}{\Delta\lambda^2-\kappa^2J^2}}.
\end{equation}
As a result, the parametric equation $\phi=\phi(\lambda)$ is
\begin{equation}
	\phi(\lambda)=\phi_0-\frac{1}{\kappa}\coth^{-1}\left(\frac{\Delta\lambda}{\kappa J}\right),
    \label{parametric_phi}
\end{equation}
where $\phi_0$ is a constant. Combining Eqs. (\ref{parametric_t}) and (\ref{parametric_phi}), one gets the equation of the trajectory
\begin{equation}
t=\pm \frac{t_0}{\left|\sinh\kappa\Delta\phi\right|},
\label{trajec_eq}
\end{equation}
where $t_0=\kappa\left|J\right|$ and $\Delta\phi=\phi-\phi_0$. The above equation is a Poinsot spiral \cite{lawrence2013catalog}, with the $+$ ($-$) sign corresponding to the upper (lower) cone. Furthermore, for $J\not= 0$ one can see that depending on the choice of the signal (or sheet of the cone), the particle always remains in the upper or lower region with no link between those regions of spacetime (see Fig. \ref{time_geodesic_fig}), but for $J=0$, which corresponds to $\phi=\phi_0$ and $t=\pm  \,\Delta\lambda$, the geodesics are straight lines. 

On the other hand, from Eqs. (\ref{tchange}) and (\ref{xchange}) one can see that all the trajectories given by Eq. (\ref{trajec_eq}) are straight lines in Minkowski spacetime,
\begin{equation}
X=\pm\frac{t_0}{\cosh \kappa\phi_0}+T\tanh \kappa\phi_0,
\label{trajec_minkowski}
\end{equation}
where for $J=0$ one recovers Eq. (\ref{chi_const}). As a result, the particle indeed can travel from one cone to the other, but such trajectories are very unstable since small perturbations in the value of $J=0$ cause large deviations on the trajectories. A similar result was found for a classical particle in a double cone in Ref. \cite{kowalski2013dynamics}, where a classical nonrelativistic particle constrained to a double cone only crosses the vertex through straight lines.

However, we are dealing with a toy model for a cyclic universe with contraction and expansion phases joined by a cosmic singularity. Thus, as pointed out in Refs. \cite{malkiewicz2006simple,poloneses2006}, due to the fact that timelike geodesics coincide with the trajectories of \textit{test} particles, which do not distort the spacetime around them, there is no obstacle for such geodesics to reach (leave) the singularity. Furthermore, if one postulates that such a particle arriving at the singularity coming from the lower cone is ``annihilated'' at the singularity, while another one is ``created'' at the upper cone, and considering that the Cauchy problem is not well defined at $t=0$, there are several types of propagation depending on the way a particle travels towards the singularity (see Sec. III of Ref. \cite{malkiewicz2006simple}). Although all of them must be consistent both with the constraint given by Eq. (\ref{geot_almost2}) and with the fact that the angular momentum $J$ is constant. It is out of the scope of this work to discuss the details and properties of such propagations.

\begin{figure}
\centering
\includegraphics[width=1.0\columnwidth]{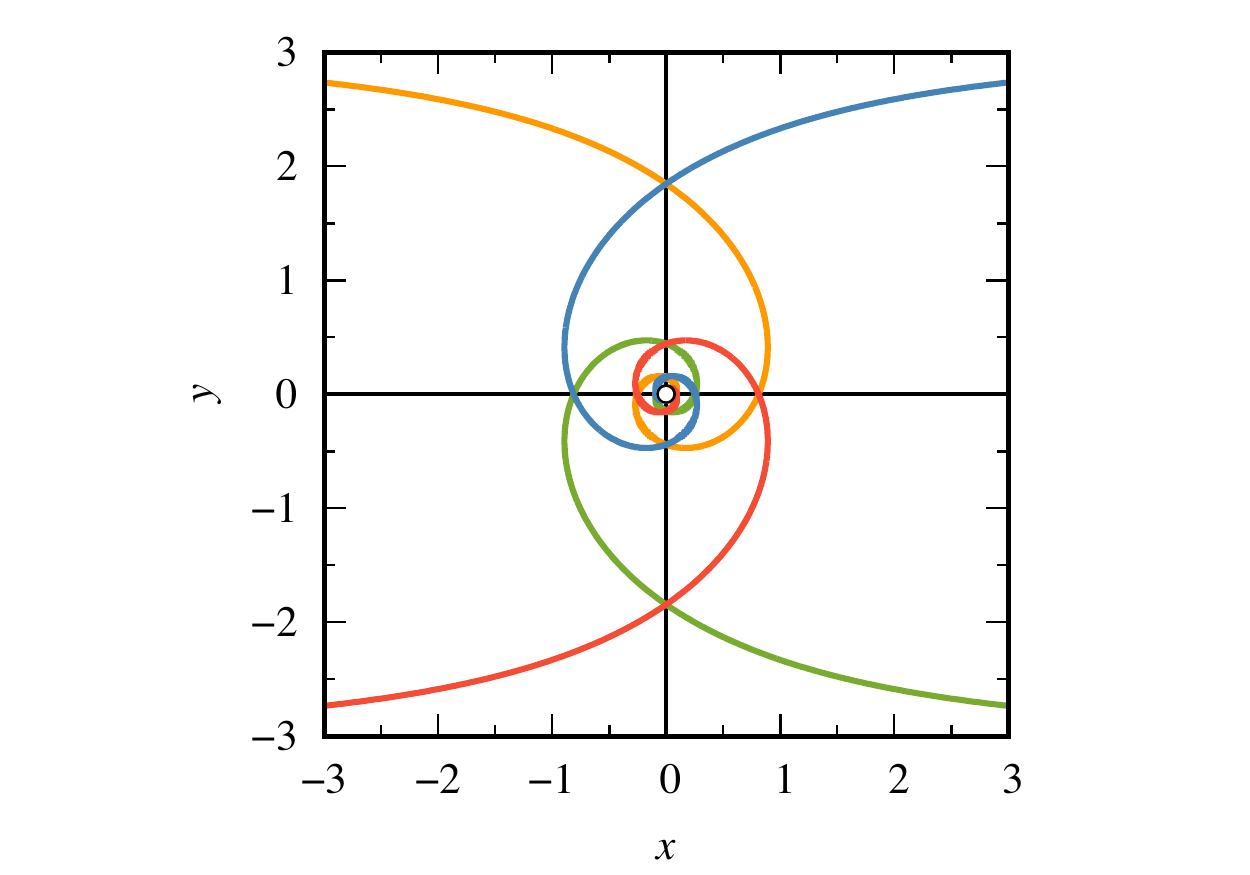}
\caption{Timelike geodesics (Poinsot spirals) with the radial time $t$ given in units of $t_0$ and $\kappa=1/3$, corresponding to the curves in the previous figure. The blue and green (orange and red) lines are moving away (towards) from (to) the singularity. Furthermore, particles following the trajectories in the first (third) and second (fourth) quadrants are spinning clockwise (counterclockwise). The blank point at the origin is just to emphasize that the curves do not reach the singularity.}
\label{poinsot_milne_fig}
\end{figure}

Next we present the timelike geodesics in a polar plot, which provides a clearer way to visualize the trajectories (see Fig. \ref{poinsot_milne_fig}). As already shown in Fig. \ref{time_geodesic_fig}, the trajectories in the lower cone are going towards  the cosmic singularity, which means that particles in the red (orange) curve have positive (negative) angular momentum $J$ and therefore are spinning in the counterclockwise (clockwise) direction. For the upper cone, particles traveling along the green (blue) curve have positive (negative) values of $J$ and are spinning counterclockwise (clockwise). As a result, in a transition from the red to the green curve the angular momentum is conserved, while for a transition to the blue one, $J$ would not be conserved. Clearly, the same reasoning applies for particles coming from the orange curve.

For the purpose of completeness, we perform the same calculations for spacelike geodesics, which can be worldlines of tachyons \cite{feinbergTachyon}. The procedure is the same, the major change being the choice of the constant of integration in Eq. (\ref{geot_almost}). Therefore, because the momentum and velocity must be spacelike, it is useful to choose $u^iu_i=-2E=1$. Thus, Eq. (\ref{geot_almost}) becomes
\begin{equation}
	\dot{t}^2-\frac{\kappa^2J^2}{t^2}=-1,
    \label{geot_almost3}
\end{equation}
the rest of the procedure being rather straightforward and leading to the following parametric equations
\begin{align}
	t(\lambda)&=\pm\sqrt{-\Delta\lambda^2+\kappa^2J^2}, \label{spacelike_t}\\
    \phi(\lambda)&=\phi_0+\frac{1}{\kappa}\tanh^{-1}\left(\frac{\Delta\lambda}{\kappa J}\right),\label{spacelike_phi}
\end{align}
where $\Delta\lambda=\lambda-\lambda_0$, with $\lambda$, $\lambda_0$  and $\phi_0$ denoting an affine parameter and constants of integration, respectively. As for the trajectory $t=t(\phi)$,
\begin{equation}
	t=\pm\frac{t_0}{\cosh \kappa\Delta\phi},
    \label{spacelike_trajec}
\end{equation}
where $t_0=\kappa\left|J\right|$ and $\Delta\phi=\phi-\phi_0$. Equation (\ref{spacelike_trajec}) is also a Poinsot spiral, with the $+$ ($-$) sign corresponding to the upper (lower) cone. Furthermore, as in the previous case the trajectories are straight lines in Minkowski spacetime,
\begin{equation}
	X=\mp \frac{t_0}{\sinh\kappa\phi_0}+T\coth\kappa\phi_0,
    \label{spacelike_minkowski}
\end{equation}
where the $-$ $(+)$ corresponds to the upper (lower) sheet.

\begin{figure}
	\centering
    \includegraphics[width=1.0\columnwidth]{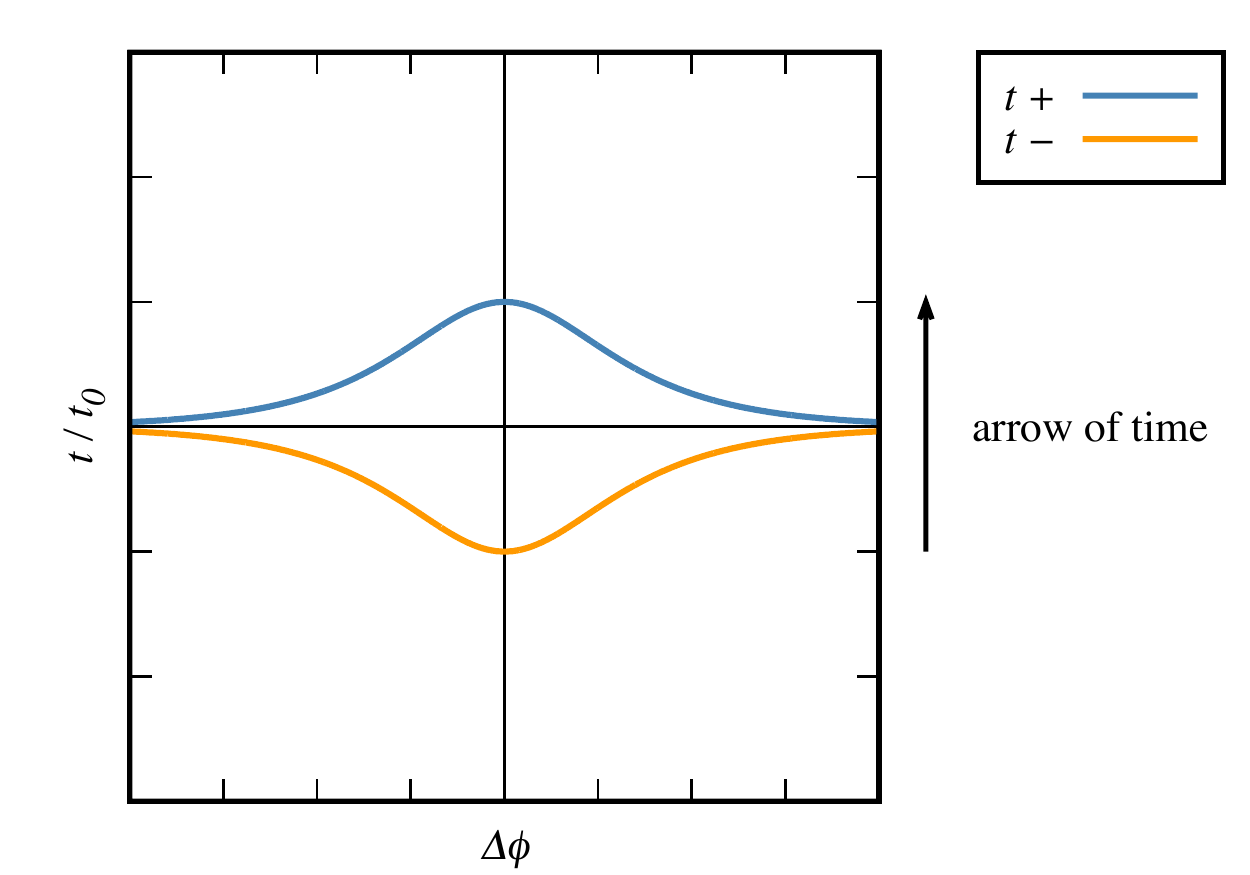}
    \caption{Graph for the geodesic equation given by Eq. (\ref{spacelike_trajec}) with $t$ in units of $t_0$, with the blue curve for the upper cone and the orange curve for the lower one. A tachyon coming from the singularity (from the left of the graph) with $J>0$ accelerates until the end of its time $t=t_0$ and then decelerates towards the singularity being ultimately annihilated. This creates a tachyon on the other sheet which goes through the same process.}
    \label{space_geodesic}
\end{figure}

\begin{figure}
	\centering
    \includegraphics[width=1.0\columnwidth]{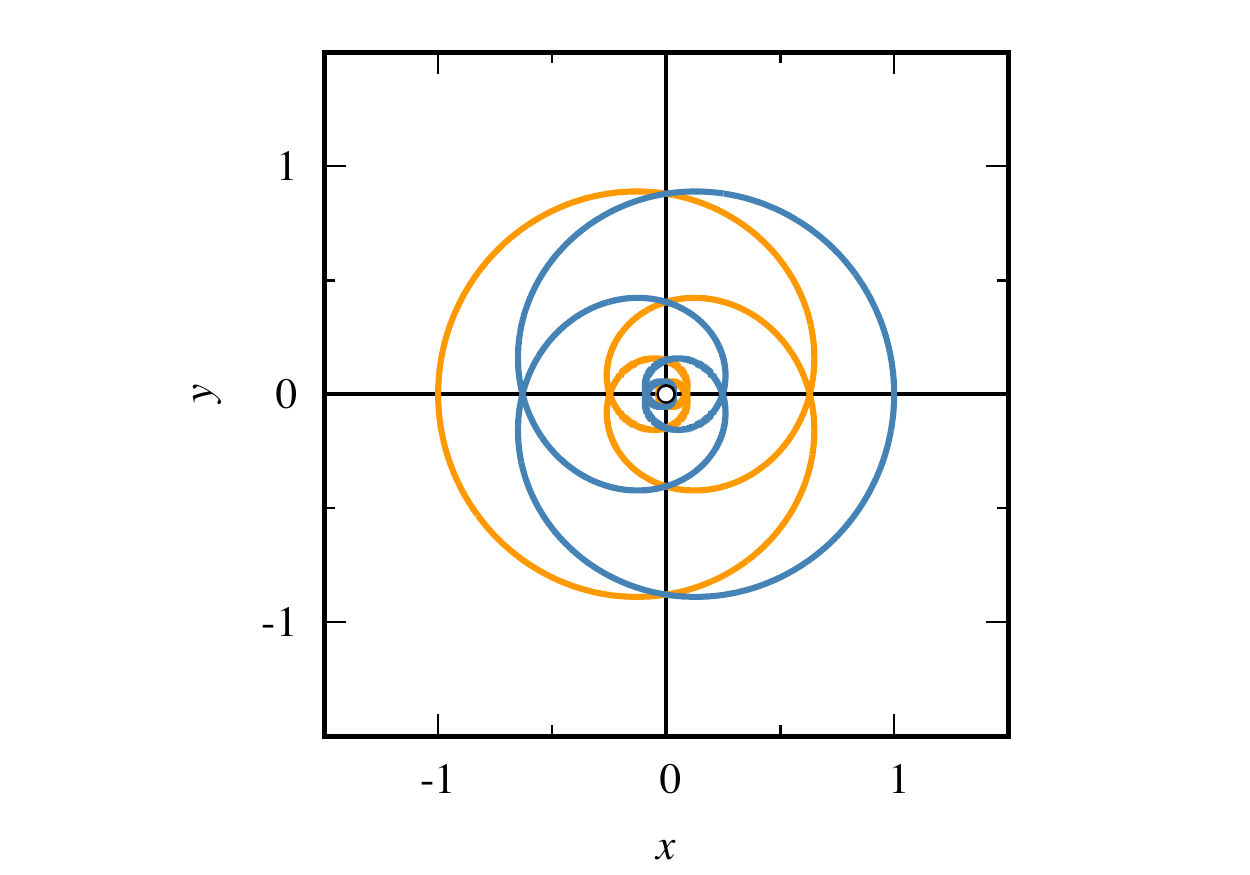}
    \caption{Spacelike geodesics (Poinsot spirals) with radial time $t$ given in units of $t_0$ and $\kappa=1/3$, corresponding to the curves in the previous figure. The closed curves show that tachyons spiral outward from the singularity and inward to the singularity on the same sheet of the cone.}
    \label{space_polar}
\end{figure}

From Eq. (\ref{spacelike_trajec}) the variable $t$ has $t_0$ as its  limiting value. To clarify this result, we remark that it was pointed out by Feinberg \cite{feinbergTachyon} that tachyons lose energy as they speed up. Thus, from Eq. (\ref{spacelike_trajec}) the velocity $\sqrt{dx^idx_i}/dt$ (with $i=1,2,3$),
\begin{equation}
	\left|\kappa t\,\frac{d\phi}{dt}\right|=\frac{t_0}{\sqrt{{t_0}^2-t^2}}
    \label{tachyon_speed}
\end{equation}
ranges from $(1,\infty)$ as $t$ ranges between 0 and $t_0$, respectively (note that for a lightlike interval, $\kappa t\,d\phi/dt=1$). Therefore, as the tachyon comes accelerating from the singularity, we can see from Eq. (\ref{geot_almost3}) that the time component of the momentum $p^0=\dot{t}$ (which is associated with the energy) decreases, reaching its minimum value at $t_0$. From that point on it starts to increase as the tachyon decelerates towards  the singularity (see Fig. \ref{space_geodesic}). This leads to an interpretation of tachyons being created and annihilated at the same sheet of the cone, which can be the upper sheet or the lower one. Every annihilation on a sheet creates a tachyon on the other sheet, in an endless cycle. This can be seen in a polar plot as closed curves in spacetime (see Fig. \ref{space_polar}).

The purpose of this section was to calculate the classical trajectories for both ordinary particles as well as tachyons through the least action principle, showing the transitions which may occur between the two cones. Even though we were dealing with classical particles, it happens as if the particles of either kind go through a process of  annihilation/creation  in order to cross the singularity. In the next section we propose a geometric optics model that emulates the trajectories of the particles in $\mathcal{M}_C$ in a hyperbolic metamaterial.
\section{A metamaterial model for the $\bm{\mathcal{M}_C}$ universe}
\label{sec4}
In order to simulate Milne particles in a condensed matter system, we  study the propagation of light in a hyperbolic liquid crystal metamaterial, HLCM, presented in Ref. \cite{lavrentovich2012liquid}.
Our study on light propagation lies in the realm of geometrical (or ray) optics, which essentially involves the application of Fermat's principle along with the variational principle that determines the path followed by light (geodesics). Therefore we seek an extremum of the integral
\begin{equation}
	S=\int_A^B{N_edl},
    \label{fermat_principle}
\end{equation}
where $dl$ is the element of arclength along the path between points $A$ and $B$, $N_e$ is the effective refractive index of the material and the product $N_e dl$ between them is called ``optical path.''\footnote{It was shown in Ref. \cite{observer_nindex} that the refractive index is dependent of the observer. Here we consider ``static observers," which are at rest with respect to the space coordinates with the time component of the velocity $u^t$ tangent to the coordinate time axis.} Because we are dealing with an anisotropic medium, there are two distinct polarizations for the light rays, namely, the ordinary and extraordinary rays. In the former, the polarization (electric field) is perpendicular to both the director $\mathbf{\hat{n}}$ (i.e. the unit vector along the average orientation of the molecular rods constituting the nematic medium) and the wave vector $\mathbf{k}$. In this case, light propagates as in an isotropic medium of refractive index $n_o$ with velocity $c/n_o$. As for the extraordinary ray, the polarization lies in the plane formed by $\mathbf{\hat{n}}$ and $\mathbf{S}$. Further, the direction of the Poynting vector $\mathbf{S}$ differs from the direction of $\mathbf{k}$, which means that the energy velocity differs from the phase velocity, leading to two different refractive indexes: the ray index $N_r$, associated to the energy velocity, and the phase index $N_p$, associated with the phase velocity \cite{modern_optics}. In this work, we  discuss only the extraordinary ray.

\begin{figure}
	\centering
    \includegraphics[width=1.0\columnwidth]{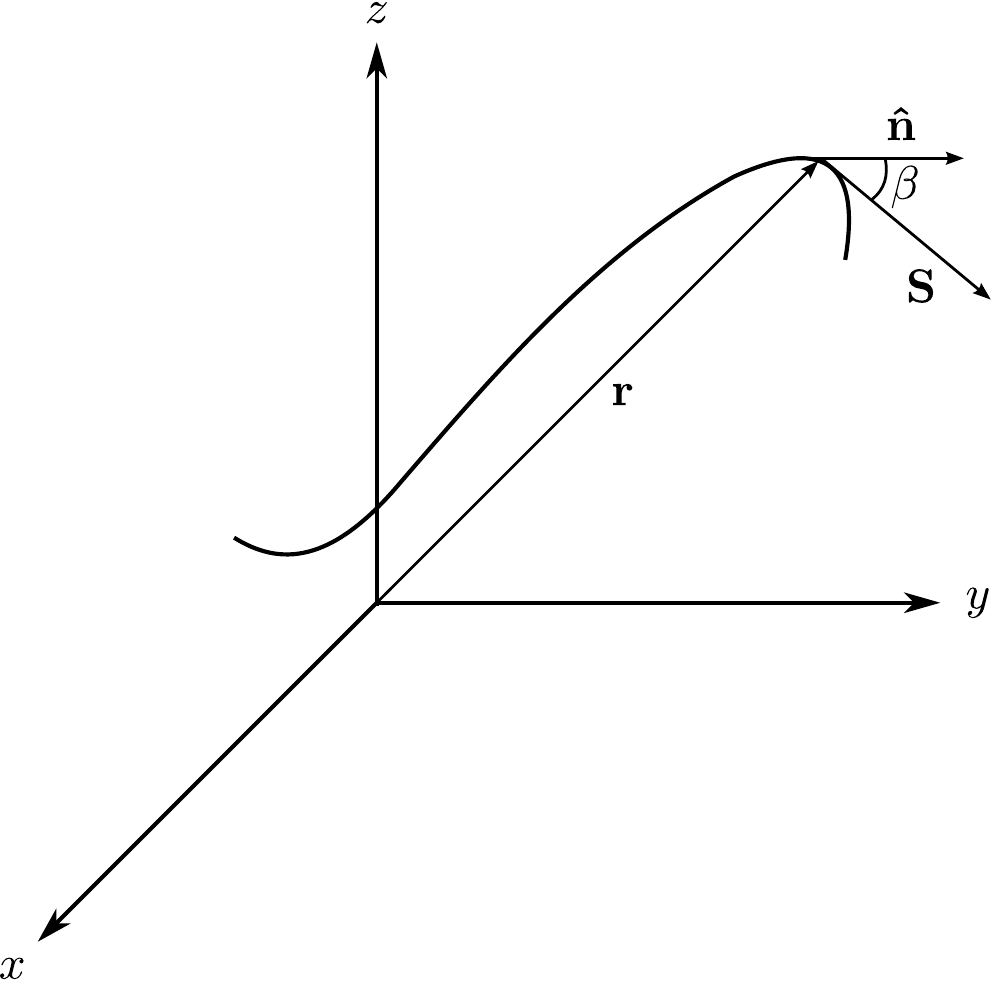}
    \caption{Curve described by the position vector $\mathbf{r}$ showing the path followed by the extraordinary light rays. The Poynting vector $\mathbf{S}$ is tangent to the curve and makes an angle of $\beta$ with the director field $\mathbf{\hat{n}}$ (optical axis).}
    \label{light_path}
\end{figure}

The application of Fermat's principle to the extraordinary light  grants us the path followed by the energy. Therefore, the effective refractive index  $N_{\text{e}}$ in Eq. (\ref{fermat_principle}) is the ray index $N_r$ and is given by \cite{maxborn_optics}
\begin{equation}
	N_r^2=n_o^2{\cos}^2\beta+n_e^2{\sin}^2\beta,
    \label{eff_refractive_index}
\end{equation}
where $\beta$ is the angle between the director $\mathbf{\hat{n}}$ and the Poynting vector $\mathbf{S}$ (see FIG. \ref{light_path}).

	Fermat's principle  essentially sums up to the geodesic determination by requiring $\delta\int_A^B\gamma_{ij}dx^idx^j=0$. Therefore, we may think of the curved trajectories of light rays in an anisotropic material as geodesics which can be found from the identification
	\begin{equation}
		N_r^2dl^2=\gamma_{ij}dx^idx^j,
        \label{link_eq}
	\end{equation}
where we introduce the  $\gamma_{ij}$ notation  to emphasize that we are dealing with a three-dimensional metric. The angle $\beta$ is determined from the specific configurations of the director field $\mathbf{\hat{n}}$ as follows. If the curve $\mathbf{r}(\lambda)$, where $\lambda$ is a parameter along the curve, represents the light trajectory, then
    \begin{equation}
    	\mathbf{t}(\lambda)=\frac{d\mathbf{r}}{d\lambda}=\frac{\mathbf{S}}{||\mathbf{S}||}
        \label{tangent}
    \end{equation}
is the tangent vector at each position parametrized by $\lambda$. If the parameter $\lambda$ is the arclength $l$, then from the theory of the differential geometry of curves \cite{willmore_diffgeom}, $||\mathbf{t}||=1$. Thus, fixing this choice $\lambda=l$
 and also using the fact that $||\mathbf{\hat{n}}||=1$,
 	\begin{equation}
 		\cos\beta=\mathbf{t}\cdot\mathbf{\hat{n}}.
        \label{cosbeta}
 	\end{equation}
  
To proceed further, one has to know the expression for the director $\mathbf{\hat{n}}$, which depends on the system in question. In what follows, we show the form for $\mathbf{\hat{n}}$ that is suitable for our cosmological analog model. The hyperbolic behavior of the metamaterial is characterized by a topological defect called disclination. In our case, the liquid crystal is in a nematic phase and such disclinations are classified according to the topological index (or strength) $k$ which gives a measure of how much the director $\mathbf{\hat{n}}$ rotates as one goes around the defect. Thus, following the same approach as Refs. \cite{satiro/moraes2006,satiro/moraes2008,erms/moraes2011}, the director configurations (in the $xy$ plane) are given by
\begin{equation}
	\vartheta(\phi)=k\phi+c,
    \label{disclin_angle}
\end{equation}
where $\vartheta$ is the angle between the molecular axis and the $x$-axis, $\phi$ is the usual azimuthal angle in cylindrical or spherical coordinates and $c$ is a constant parameter. In the case discussed here, the disclinations are such that the system presents a translational symmetry along the $z$-axis. Therefore, we have effectively a two-dimensional problem and the director $\mathbf{\hat{n}}$ is given in Cartesian coordinates by \cite{satiro/moraes2006,satiro/moraes2008,erms/moraes2011}
	\begin{equation}
		\mathbf{\hat{n}}=(\cos\vartheta,\sin\vartheta,0).
        \label{director}
	\end{equation}
   
\begin{figure}
	\centering
    \includegraphics[width=.6\columnwidth]{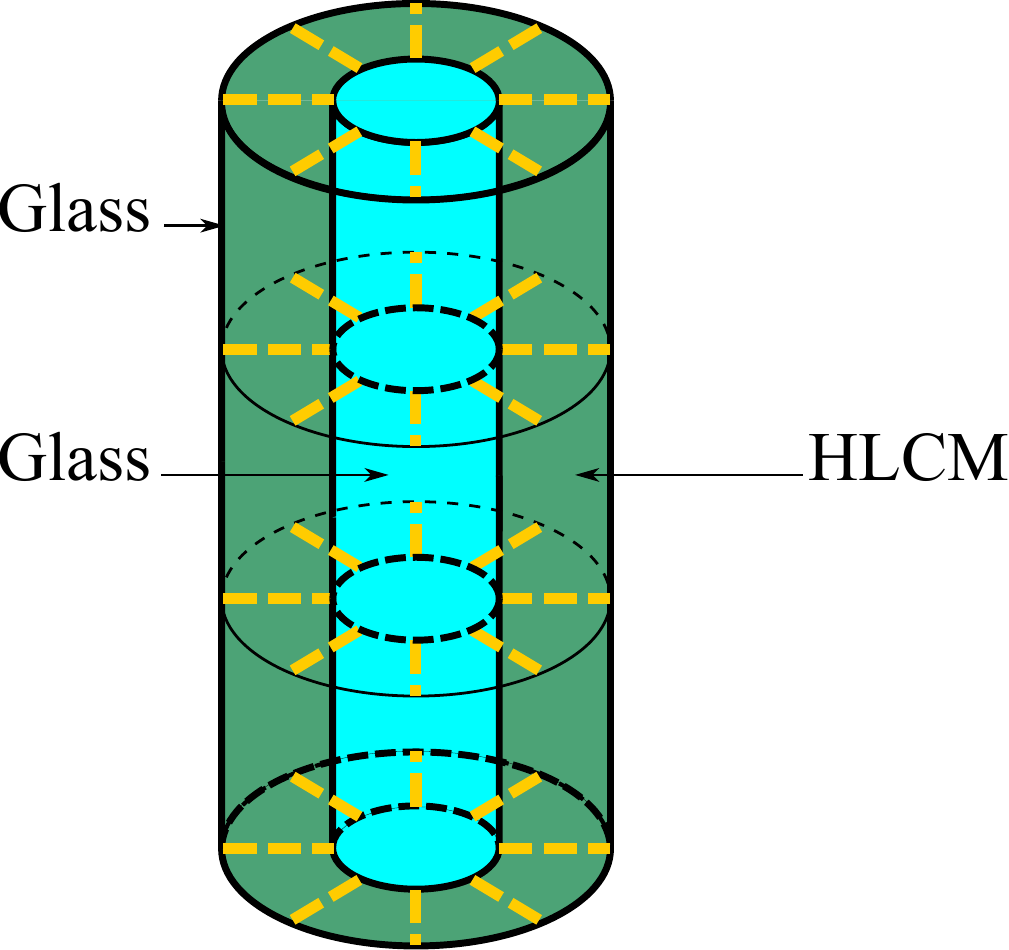}
    \caption{Director configurations for HLCM in a cylindrical shell, with optical axis represented in cylindrical coordinates $(\rho,\phi,z)$ as $\mathbf{\hat{n}}=\bm{\hat{\rho}}$. The walls of the inner and outer cylinder are made of glass, while the liquid crystal together with metallic nanorods (aligned in the direction of $\mathbf{\hat{n}}$) fill the space in between.}
    \label{fig1}
\end{figure}

In order to simulate Klein-Gordon (KG) particles in a Milne universe, let us consider a radial director field  (see Fig. \ref{fig1})
	\begin{equation}
		\mathbf{\hat{n}}=(\cos\phi,\sin\phi,0),
        \label{radial_director}
	\end{equation}
where in this case $k=1$ and $c=0$ in Eq. (\ref{disclin_angle}). For convenience we will use cylindrical coordinates $(\rho,\phi,z)$, where $\mathbf{r}(l)=\rho\bm{\hat{\rho}}+z\bm{\hat{z}}$ and $\mathbf{\hat{n}}=\bm{\hat{\rho}}$ by Eq. (\ref{radial_director}). Therefore, the tangent vector $\mathbf{t}$ will be
	\begin{equation}
		\mathbf{t}=\frac{d\mathbf{r}}{dl}=\dot{\rho}\bm{\hat{\rho}}+\rho\dot{\phi}\bm{\hat{\phi}}+\dot{z}\bm{\hat{z}},
        \label{tangent_radial}
	\end{equation}
    where the dot stands for $d/dl$. Then, from Eq. (\ref{cosbeta}) we have
    \begin{equation}
    	\cos\beta=\dot{\rho}.
        \label{cosbeta2}
    \end{equation}
    Next, let us consider the Euclidean line element
	\begin{equation}
		dl^2=d\rho^2+\rho^2d\phi^2+dz^2,
        \label{dl_cylindrical}
	\end{equation}
which leads to the following relation
	\begin{equation}
		\dot{\rho}^2+\rho^2\dot{\phi}^2+\dot{z}^2=1.
        \label{dl/dl_cylindrical}
	\end{equation}
    Therefore, combining Eqs. (\ref{cosbeta2}) and (\ref{dl/dl_cylindrical}) one gets
    \begin{equation}
    	\sin\beta=\sqrt{\rho^2\dot{\phi}^2+\dot{z}^2}.
        \label{sinbeta}
    \end{equation}
    
The ray index $N_r$ can now be obtained with the help of Eqs. (\ref{eff_refractive_index}), (\ref{cosbeta2}) and (\ref{sinbeta}) as
    \begin{equation}
    	N_r^2=n_o^2\dot{\rho}^2+n_e^2(\rho^2\dot{\phi}^2+\dot{z}^2).
        \label{ray_index}
    \end{equation}
Then, the line element $ds^2=N_r^2dl^2=\gamma_{ij}dx^idx^j$ will have the following form
    \begin{equation}
    	ds^2=n_o^2d\rho^2+n_e^2\rho^2d\phi^2+n_e^2dz^2.
        \label{effect_metric}
    \end{equation}
    
Because we are dealing with metamaterials, it is more useful to write the refractive indexes $n_o$ and $n_e$ as functions of the components of the permittivity tensor $\epsilon_{ij}$ of the material. Therefore, introducing the usual notation, as in Ref. \cite{kleman2007soft}, where $n_o^2=\epsilon_{\perp}$ and $n_e^2=\epsilon_{\parallel}$, Eq. (\ref{effect_metric}) becomes
\begin{equation}	ds^2=\epsilon_{\perp}d\rho^2+\epsilon_{\parallel}\rho^2d\phi^2+\epsilon_{\parallel}dz^2,
\label{effect_metric2}
\end{equation}
where in this case $\epsilon_{\perp}=\epsilon_{\phi\phi}=\epsilon_{zz}$ and $\epsilon_{\parallel}=\epsilon_{\rho\rho}$. Furthermore, the permittivity tensor $\bm{\epsilon}$ is given by \cite{modern_optics,maxborn_optics}
\begin{equation}	\bm{\epsilon}=\epsilon_{\parallel}\bm{\hat{\rho}}\otimes\bm{\hat{\rho}}+\epsilon_{\perp}\bm{\hat{\phi}}\otimes\bm{\hat{\phi}}+\epsilon_{\perp}\bm{\hat{z}}\otimes\bm{\hat{z}}.
\label{epsilon_tensor}
\end{equation}
As pointed out before, the system in question has a translational symmetry along the $z$-axis, and therefore we are interested in the extraordinary ray propagating in $z=\text{constant}$ planes. Thus, $dz=0$ in Eq. (\ref{effect_metric2}). Furthermore, the metric tensor $\gamma_{ij}$ describes a ``real'' three-dimensional space, having only spatial coordinates. However, light propagates along null geodesics in a four-dimensional spacetime. Therefore, we use the fact that the application of Fermat's principle in a three-dimensional metric with only spatial coordinates is equivalent to calculating null geodesics in a four-dimensional spacetime with a pseudo-Riemannian metric $g_{ij}$. Thus, the relation between the metrics $\gamma_{ij}$ and $g_{ij}$ is as follows (see p. 1108 of Ref. \cite{misner1973gravitation})
\begin{equation}
	\gamma_{ij}=-\frac{g_{ij}}{g_{00}}.
    \label{metric_correlation}
\end{equation}
Taking $g_{00}=1$, Eq. (\ref{effect_metric2}) becomes
\begin{equation}
	ds^2=dT^2-\epsilon_{\perp}d\rho^2-\epsilon_{\parallel}\rho^2d\phi^2,
    \label{effect_metric3}
\end{equation}
where $T$ is the Minkowskian time.

In what follows, we explore the property of metamaterials to produce negative permittivities. Due to the fact the director field is radial ($\mathbf{\hat{n}}=\bm{\hat{\rho}}$) and the metallic nanorods are aligned in the same direction, we have a metallic behavior along the radial coordinate and can therefore obtain $\epsilon_{\parallel}<0$. Furthermore, considering $\epsilon_{\perp}>0$ and rescaling the radial coordinate by $r=\rho\sqrt{\epsilon_{\perp}}$ one gets
\begin{equation}
	ds^2=dT^2-dr^2+\alpha^2r^2d\phi^2,
    \label{disclination_metric}
  \end{equation}
where $\alpha^2=|\epsilon_{\parallel}|/\epsilon_{\perp}$ is the disclination parameter associated to a hyperbolic (imaginary) deficit angle of $i\times2\pi\alpha$ \cite{fumeron2015optics}. Also note that  the spatial part of the metric above is equivalent to the $\mathcal{M}_C$ metric given by Eq. (\ref{milne_metric}), with $r$ behaving as the timelike variable $T$.

By the same procedure developed in the previous section, Eq. (\ref{disclination_metric}) gives the Lagrangian
\begin{equation}
	\mathcal{L}_\text{kin}=\frac{\dot{T}^2}{2}-\frac{\dot{r}^2}{2}+\frac{\alpha^2r^2\dot{\phi}^2}{2},
    \label{disclin_lagrangian}
\end{equation}
which leads to the following geodesic equations
\begin{align}
	\frac{dT}{d\lambda}&=\nu,\label{geodisc_time}\\
    \ddot{\phi}+\frac{2}{r}\,\dot{r}\dot{\phi}&=0,\label{geodisc_phi}\\
    \ddot{r}+\alpha^2r\dot{\phi}^2&=0\label{geodisc_r},
\end{align}
where $\lambda$ is an affine parameter and $\nu$ is a constant. The first equation just states the conservation of the canonical momentum $p_T$, while the other two are equivalent to the geodesic equations for classical particles in $\mathcal{M}_C$ given by Eqs. (\ref{geo_eq_p}) and (\ref{geo_eq_t}). It is possible to follow the same steps to integrate the equations as we did in Sec. \ref{classical_MC}. However, as in this case we are dealing with null geodesics, $g_{ij}\dot{x}^i\dot{x}^j=0$ and therefore one gets the following useful constraint
\begin{equation}
	\dot{r}^2-\frac{\alpha^2L^2}{r^2}=\nu^2,
	\label{null_geodesic_constraint}
\end{equation}
with the constant $L=r^2\dot{\phi}$ being the angular momentum. The previous equation is essentially the same as Eq. (\ref{geot_almost2}). The geodesic equations for this case were already solved in Ref. \cite{fumeron2015optics} and the solution for the trajectory is the following
\begin{equation}
	r=\frac{\alpha\,r_0}{|\sinh\alpha\Delta\phi|},
	\label{trajectory_disclination}
\end{equation}
where $r_0=|L|/\nu$. As a result, the path followed by light rays in the three-dimensional space in the metamaterial are Poinsot's spirals like the timelike geodesics in $\mathcal{M}_C$ found in Sec. \ref{classical_MC}. Comparing Eqs. (\ref{trajec_eq}) and (\ref{trajectory_disclination}), the radial coordinate $r$ behaves like the time coordinate $t$ and $\alpha$ is the analogue of the constant $\kappa$. As pointed out in Ref. \cite{fumeron2015optics}, the hyperbolic behavior of the material generates a force towards the defect and the parameter $\alpha$ is directly related to it, with $1/\alpha$ being understood as the vorticity of the defect (the smaller the value of $\alpha$, the stronger is the attraction towards the defect). Furthermore, since the defect generates an attraction, the trajectories given by Eq. (\ref{trajectory_disclination}) can be interpreted as the equivalent to geodesics in $\mathcal{M}_C$ going to the cosmic singularity (big crunch).
\begin{figure}
	\centering
    \includegraphics[width=.6\columnwidth]{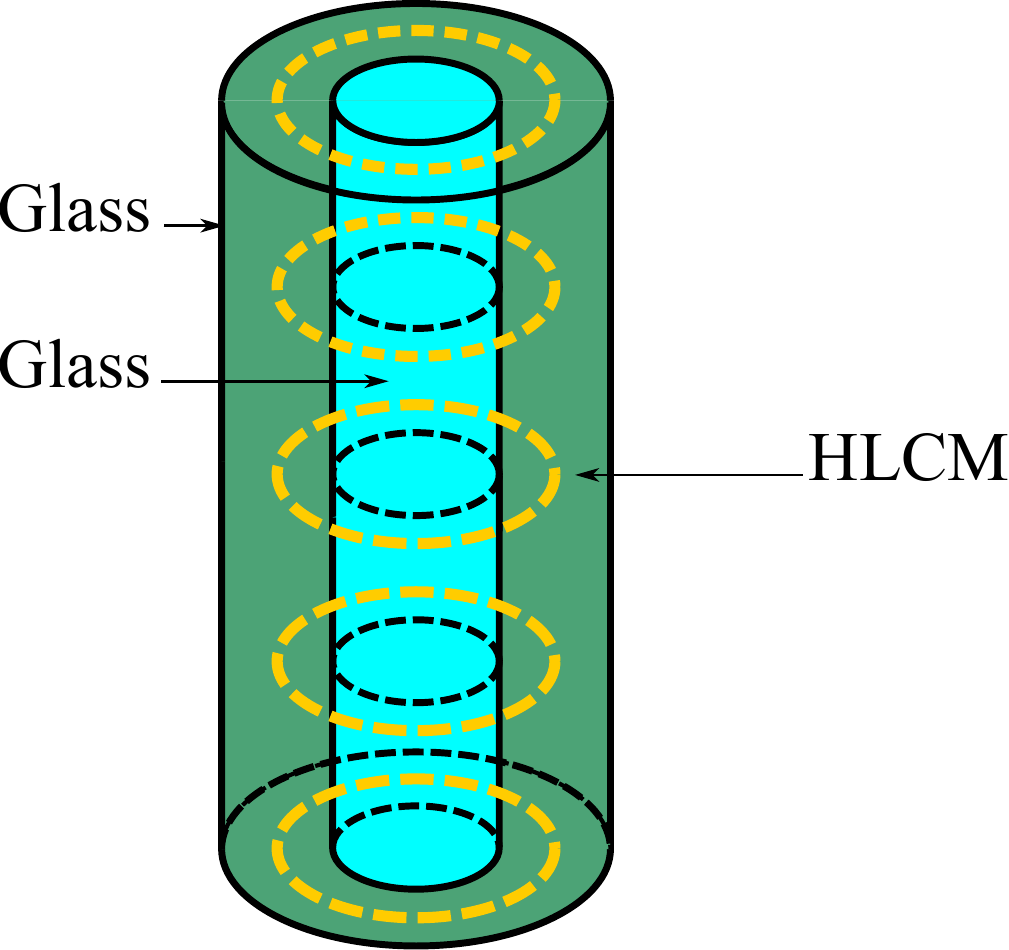}
    \caption{Director configurations for HLCM in a cylindrical shell for a circular field, with optical axis represented in cylindrical coordinates $(\rho,\phi,z)$ as $\mathbf{\hat{n}}=\bm{\hat{\phi}}$.}
    \label{fig2}
\end{figure}

To simulate tachyons, we consider the director field as $\mathbf{\hat{n}}=\bm{\hat{\phi}}$ (see Fig. \ref{fig2}), meaning that $k=1$ and $c=\pi/2$ in Eq. (\ref{director}). In this case, our four-dimensional line element will be
\begin{equation}
	ds^2=dT^2-\epsilon_{\parallel}d\rho^2-\epsilon_{\perp}\rho^2d\phi^2,
    \label{hlcm_metric_tachyon}
\end{equation}
where now $\epsilon_{\parallel}=\epsilon_{\phi\phi}$ and $\epsilon_{\perp}=\epsilon_{\rho\rho}$. Thus, considering $\epsilon_{\parallel}<0$, $\epsilon_{\perp}>0$ and rescaling the radial coordinate by $r=\rho\sqrt{|\epsilon_{\parallel}|}$ one gets
\begin{equation}
	ds^2=dT^2+dr^2-\alpha^2r^2d\phi^2,
    \label{hlcm_metric_tachyon2}
\end{equation}
where $\alpha^2=\epsilon_{\perp}/|\epsilon_{\parallel}|$. Therefore, the Lagrangian will be
\begin{equation}
	\mathcal{L}_{\text{kin}}=\frac{\dot{T}^2}{2}+\frac{\dot{r}^2}{2}-\frac{\alpha^2r^2\dot{\phi}^2}{2},
    \label{hlcm_tachyon_lagrangian}
\end{equation}
which also gives Eqs. (\ref{geodisc_time})--(\ref{geodisc_r}) as geodesic equations. However, as in Sec. \ref{classical_MC} we will have a different constraint. Thus, for $g_{ij}\dot{x}^i\dot{x}^j=0$,
\begin{equation}
	\dot{r}^2-\frac{\alpha^2L^2}{r^2}=-\nu^2,
    \label{null_geo_constraint2}
\end{equation}
which is the analogue of Eq. (\ref{geot_almost3}). By the same procedure done in the previous cases, the trajectory will be
\begin{equation}
	r=\frac{\alpha\,r_0}{\cosh\alpha\Delta\phi},
    \label{trajectory_disclination2}
\end{equation}
where $r_0=|L|/\nu$ again. Therefore, we obtain the analogue of the spacelike geodesics in $\mathcal{M}_C$ found in Sec. \ref{classical_MC}.

We remark that in Ref. \cite{semiclassical_hyperlens} both trajectories given by Eqs. (\ref{trajectory_disclination}) and (\ref{trajectory_disclination2}) were obtained through numerical simulations in a scattering experiment in a hyperlens system. This is interesting to represent and visualize the double cone geometry; since light going to the disclination can be viewed as trajectories in the lower cone and light coming out of the disclination as the trajectories in the upper one.

Having shown that the classical behavior can be emulated, in the following section it will be shown that the quantum behavior of KG particles in $\mathcal{M}_C$ also has an analogy in the framework of wave optics.
\section{Mimicking quantum particles in $\bm{{\mathcal{M}_C}}$}
In this last section, our goal is to show how to mimic a KG particle in $\mathcal{M}_C$ through the system presented in the previous section. Therefore, let us consider the KG equation for a scalar field $\varphi$ (in natural units $c,\hbar=1$)
\begin{equation}
	\left(\Delta-m^2\right)\varphi=0,
    \label{CM_KG}
\end{equation}
where $\Delta$ is the Laplace-Beltrami operator as presented in Sec. \ref{CM_section}. For the $\mathcal{M}_C$ universe, $\Delta_{\mathcal{M}_C}$ was given in Eq. (\ref{laplacian_CM}). Thus,
\begin{equation}
	\left[-\frac{1}{t}\frac{\partial}{\partial t}\left(t\frac{\partial}{\partial t}\right)+\frac{1}{\kappa^2t^2}\frac{\partial^2}{\partial\phi^2}-m^2\right]\varphi=0.
    \label{CM_KG2}
\end{equation}


Concerning the metamaterial model, we examine the propagation of light in the scalar wave approximation, allowing the use of the covariant d'Alembert wave equation
\begin{equation}
	\nabla_i\nabla^i\Phi=\frac{1}{\sqrt{-g}}\partial_i\left(
    \sqrt{-g}g^{ij}\partial_j\Phi\right)=0,
    \label{dalembert_weq}
\end{equation}
where $\Phi$ is the wave function. Therefore, according to the metric given by Eq. (\ref{effect_metric3}), one gets
\begin{equation}
	-\frac{1}{\epsilon_{\perp}}\frac{1}{\rho}\frac{\partial}{\partial\rho}\left(\rho\frac{\partial\Phi}{\partial\rho}\right)-\frac{1}{\epsilon_{\parallel}}\frac{1}{\rho^2}\frac{\partial^2\Phi}{\partial\phi^2}+\frac{\partial^2\Phi}{\partial T^2}=0.
    \label{dalembert_eq2}
\end{equation}
Assuming a harmonic dependence in time, we make a separation of variables $\Phi(t,\rho,\phi)=e^{-i\omega T}\psi(\rho,\phi)$. Thus,
\begin{equation}
	-\frac{1}{\epsilon_{\perp}}\frac{1}{\rho}\frac{\partial}{\partial\rho}\left(\rho\frac{\partial\psi}{\partial\rho}\right)-\frac{1}{\epsilon_{\parallel}}\frac{1}{\rho^2}\frac{\partial^2\psi}{\partial\phi^2}-\omega^2\psi=0,
    \label{almost_KG}
\end{equation}
where $\omega$ is the propagation frequency of the light or effective mass of an analogue KG particle. Furthermore, recalling our previous choice of $\epsilon_{\perp}=\epsilon_{\phi\phi}>0$ and $\epsilon_{\parallel}=\epsilon_{\rho\rho}<0$, one gets
\begin{equation}
	\left[-\frac{1}{\epsilon_{\perp}}\frac{1}{\rho}\frac{\partial}{\partial\rho}\left(\rho\frac{\partial}{\partial\rho}\right)+\frac{1}{|\epsilon_{\parallel}|}\frac{1}{\rho^2}\frac{\partial^2}{\partial\phi^2}-\omega^2\right]\psi=0.
    \label{almost_KG2}
\end{equation}
The equation above is similar to Eq. (\ref{CM_KG2}) and it was already used to mimic KG particles in plasmonic metamaterials \cite{smolyaninov2011modeling,smolyaninov2012experimental}. In terms of $r=\rho\sqrt{\epsilon_{\perp}}$ and the parameter $\alpha^2=|\epsilon_{\parallel}|/\epsilon_{\perp}$, it becomes
\begin{equation}
	\left[-\frac{1}{r}\frac{\partial}{\partial r}\left(r\frac{\partial}{\partial r}\right)+\frac{1}{\alpha^2}\frac{1}{r^2}\frac{\partial^2}{\partial\phi^2}-\omega^2\right]\psi=0,
    \label{eq39}
\end{equation}
where $\omega$ is treated as an effective mass (in natural units). Furthermore, there is a contribution $\alpha^2$ from the disclination, which by Eq. (\ref{laplacian_Cone}) is associated to a cone angle $\theta=\sin^{-1}\alpha$, whereas in Eq.(\ref{CM_KG2}) $\theta=\sinh^{-1}\kappa$.

To solve Eq. (\ref{eq39}), we remark that the hyperbolic metamaterial obeys the dispersion relation
\begin{equation}
	\frac{k_\rho^2}{\epsilon_{\perp}}-\frac{k_{\phi}^2}{|\epsilon_{\parallel}|}=\omega^2
    \label{dispersion1}
\end{equation}
and the angular momentum conservation $l=k_{\phi}\rho$, as shown in Ref. \cite{hyperlens2006}. In terms of the angular momentum quantum number $l$ and the radial variable $r$, Eq. (\ref{dispersion1}) becomes
\begin{equation}
	k_r^2-\frac{l^2}{\alpha^2r^2}-\omega^2=0,
    \label{dispersion2}
\end{equation}
where we have used the fact that $k_{r}=k_{\rho}/\sqrt{\epsilon_{\perp}}$ due to the change of scale passing from $\rho$ to $r$. Equation (\ref{dispersion2}) is consistent with Eq. (\ref{eq39}) since, in terms of operators, it takes the following form
\begin{equation}
	\left[\hat{k}_r^2-\frac{\hat{L}_z^2}{\alpha^2r^2}-\omega^2\right]\psi=0,
    \label{pde_operator}
\end{equation}
where $\hat{k}_r^2$ and $\hat{L}_z^2$ are the usual operators
\begin{align}
	\hat{k}_r^2&=-\frac{1}{r}\frac{\partial}{\partial r}\left(r\frac{\partial}{\partial r}\right),\label{kr_op}\\
    \hat{L}_z^2&=-\frac{\partial^2}{\partial\phi^2}.\label{Lz_op}
\end{align}
Therefore, separating the variables $\psi(r,\phi)=f(r)g(\phi)$ leads to the equations
\begin{equation}
	\frac{d^2g}{d\phi^2}+l^2g=0,
    \label{ode_phi}
\end{equation}
and
\begin{equation}
	r^2\frac{d^2f}{dr^2}+r\frac{df}{dr}+\left(\omega^2r^2+\frac{l^2}{\alpha^2}\right)f=0.
    \label{ode_radial}
\end{equation}
Thus, the solution for Eq. (\ref{ode_phi}) is
\begin{equation}
	g(\phi)=Ae^{il\phi}+Be^{-il\phi},
    \label{phi_sol}
\end{equation}
where $A,\,B$ are constants of integration. As for Eq. (\ref{ode_radial}), it is a Bessel differential equation of imaginary order $il/\alpha$ with solutions \cite{olver2010nist,dunster1990bessel}
\begin{equation}
	f(r)=C\tilde{J}_{l/\alpha}(\omega r)+D\,\tilde{Y}_{l/\alpha}(\omega r),
    \label{radial_sol}
\end{equation}
which $C,\,D$ being constants of integration and $\tilde{J}_{l/\alpha},\,\tilde{Y}_{l/\alpha}$ are linearly independent solutions defined in Ref. \cite{olver2010nist} as
\begin{align}
	\tilde{J}_{l/\alpha}&=\text{sech}\left(\frac{\pi l}{2\alpha}\right)\text{Re}\,J_{il/\alpha}(\omega r),\label{NIST_eq1}\\
    \tilde{Y}_{l/\alpha}&=\text{sech}\left(\frac{\pi l}{2\alpha}\right)\text{Re}\,Y_{il/\alpha}(\omega r),\label{NIST_eq2}
\end{align}
where $\text{Re}\,J_{il/\alpha}$ and $\text{Re}\,Y_{il/\alpha}$ are the real parts of Bessel and Neumann functions, respectively. Following the same procedure for Eq. (\ref{CM_KG2}), one gets the same solutions for the scalar field $\varphi$ (as obtained in Ref. \cite{poloneses2006}) with the variable $r$ replaced by $t$ and the constants $\omega,\,\alpha$ interchanged by $m,\,\kappa$, respectively.

An interesting feature \cite{olver2010nist,dunster1990bessel} of Eqs. (\ref{NIST_eq1}) and (\ref{NIST_eq2}) is that they oscillate rapidly near the origin, as one can see from its behavior as $r\rightarrow 0^+$
\begin{align}
	\tilde{J}_{l/\alpha}(\omega r)=&\left(\frac{\tanh(\pi l/2\alpha)}{\pi l/2\alpha}\right)^{1/2}\cos\left[\frac{l}{\alpha}\ln\left(\frac{\omega r}{2}\right)-\gamma_{l/\alpha}\right]\nonumber\\
    &+O(\omega^2r^2),\label{NIST_near1}\\
    \tilde{Y}_{l/\alpha}(\omega r)=&\left(\frac{\coth(\pi l/2\alpha)}{\pi l/2\alpha}\right)^{1/2}\sin\left[\frac{l}{\alpha}\ln\left(\frac{\omega r}{2}\right)-\gamma_{l/\alpha}\right]\nonumber \\
     &+O(\omega^2r^2),\label{NIST_near2}
\end{align}
where $\gamma_{l/\alpha}$ is a constant defined as $\gamma_{l/\alpha}\equiv\arg\,\Gamma(1+il/\alpha)$, with $\Gamma$ being the Gamma function. The rapid oscillations are due to the logarithmic argument of the trigonometric functions. Also, for a fixed $l$, the smaller the value of $\alpha$, the stronger the oscillations become (reducing the value of $\alpha$ ``squeezes'' the period of the trigonometric functions). This analogous of the classical behavior in Sec. \ref{sec4} where $1/\alpha$ is related to the vorticity.

Furthermore, for $l\neq 0$ Eqs. (\ref{NIST_eq1}) and (\ref{NIST_eq2}) are not continuous through the origin, which is similar to the classical geodesics that in general do not cross the singularity. In Ref. \cite{poloneses2006} it was given an interesting interpretation for this fact constructing a Hilbert space $\mathcal{H}=\mathcal{H}^{(-)}\oplus\mathcal{H}^{(+)}$ as a direct sum of two Hilbert spaces $\mathcal{H}^{(-)}$, $\mathcal{H}^{(+)}$. The elements of $\mathcal{H}^{(-)}$ are solutions of Eq. (\ref{CM_KG2}) in the presingularity era $(t<0)$ while the ones of $\mathcal{H}^{(+)}$ are solutions in the postsingularity era $(t>0)$. That is, $\mathcal{H}=\mathcal{H}^-\oplus\mathcal{H}^+$ is a vector space whose elements $\varphi$ have the following form \cite{prugovecki1982quantum}
\begin{equation}
	\varphi=\left(\varphi^{(-)},\varphi^{(+)}\right)\in\mathcal{H}^{(-)}\times\mathcal{H}^{(+)},
\end{equation}
with an inner product
\begin{equation}
	\left \langle \varphi | \psi \right \rangle=\left \langle \varphi^{(-)} \big | \psi^{(-)} \right \rangle+\left \langle \varphi^{(+)} \big | \psi^{(+)} \right \rangle.
    \label{inner_product}
\end{equation}
Hence, vectors like $(\varphi^{(-)},0)$ and $(0,\varphi^{(+)})$ describe states of annihilation and creation of particles at the singularity $t=0$, respectively. By Eq. (\ref{inner_product}) an inner product between those kinds of states yields zero, which means no correlation between them. This reflects the loss of phase of the wave function due to the strong oscillations around the singularity.

The  $l=0$ case is rather straightforward; from Eqs. (\ref{phi_sol}), (\ref{NIST_eq1}) and (\ref{NIST_eq2}), $g(\phi)$ is constant and the Bessel functions of imaginary order reduce to the usual ones $J_0(\omega r),\,Y_0(\omega r)$. Moreover, due to the fact that in the classical case the geodesics are straight lines crossing the singularity and knowing that $Y_0$ diverges at the origin, a more satisfactory physical solution is given by $J_0$, as it is continuous and well defined at the origin (singularity) $r=0$ $(t=0)$.

Next, we consider briefly the case of tachyons. Therefore, as pointed in Ref. \cite{feinbergTachyon} tachyons can be regarded as having an imaginary ``rest'' mass $m=i\mu$, with $\mu\in\mathbb{R}$. As a result, Eq. (\ref{CM_KG2}) becomes
\begin{equation}
	\left[-\frac{1}{t}\frac{\partial}{\partial t}\left(t\frac{\partial}{\partial t}\right)+\frac{1}{\kappa^2t^2}\frac{\partial^2}{\partial\phi^2}+\mu^2\right]\varphi=0.
    \label{CM_KG3}
\end{equation}

To mimic tachyons in the metamaterial we consider again the case of a circular director field $\mathbf{\hat{n}}=\bm{\hat{\phi}}$ as in the previous section. Therefore, substituting the metric (\ref{hlcm_metric_tachyon2}) in the covariant d'Alembert wave equation (\ref{dalembert_weq}) and considering $\Phi(t,r,\phi)=e^{-i\omega T}\psi(r,\phi)$, one gets
\begin{equation}
		\left[-\frac{1}{r}\frac{\partial}{\partial r}\left(r\frac{\partial}{\partial r}\right)+\frac{1}{\alpha^2}\frac{1}{r^2}\frac{\partial^2}{\partial\phi^2}+\omega^2\right]\psi=0,
        \label{hlcm_tachyon_KG}
\end{equation}
or, in the operator form
\begin{equation}
	\left[\hat{k}_r^2-\frac{\hat{L}_z^2}{\alpha^2r^2}+\omega^2\right]\psi=0,
    \label{pde_operator2}
\end{equation}
where $\alpha^2=\epsilon_{\perp}/|\epsilon_{\parallel}|=\epsilon_{\rho\rho}/|\epsilon_{\phi\phi}|$ and $\omega$ is the analogue of $\text{Im}\,m=\mu$. The dispersion relation in this case takes the form
\begin{equation}
	-\frac{k_\rho^2}{|\epsilon_{\parallel}|}+\frac{k_{\phi}^2}{\epsilon_{\perp}}=\omega^2.
    \label{dispersion3}
\end{equation}
Hence, in terms of the angular momentum $l$, the constant $\alpha$ and $k_r=k_\rho/\sqrt{|\epsilon_{\parallel}|}$, Eq. (\ref{dispersion3}) becomes
\begin{equation}
	k_r^2-\frac{l^2}{\alpha^2r^2}+\omega^2=0,
    \label{dispersion4}
\end{equation}
which is consistent with Eq. (\ref{pde_operator2}).

Thus, separating the variables as $\psi(r,\phi)=f(r)g(\phi)$ one gets the same solution given by Eq. (\ref{phi_sol}) for $g(\phi)$. As for the radial part,
\begin{equation}
	r^2\frac{d^2f}{dr^2}+r\frac{df}{dr}-\left(\omega^2r^2-\frac{l^2}{\alpha^2}\right)f=0,
    \label{tachyon_ode_radial}
\end{equation}
which is the modified Bessel differential equation with imaginary order $il/\alpha$. The solution for this case will be
\begin{equation}
	f(r)=C\tilde{I}_{l/\alpha}(\omega r)+D\tilde{K}_{l/\alpha}(\omega r),
    \label{tachyon_KG_solution}
\end{equation}
where we kept the notation of Ref. \cite{olver2010nist}, with $\tilde{I}_{l/\alpha}=\text{Re}\,I_{il/\alpha}$ and $\tilde{K}_{l/\alpha}=K_{il/\alpha}$. The functions $I_{il/\alpha},\,K_{il/\alpha}$ are the modified Bessel functions of first and second kind, respectively. As in the previous case, their behavior near the origin is characterized by rapid oscillations. However, their asymptotic behavior is exponential \cite{olver2010nist,dunster1990bessel},
\begin{align}
	\tilde{I}_{l/\alpha}(\omega r)&=\left(\frac{1}{2\pi\omega r}\right)^{1/2}e^{\omega r}\left[1+O\left(\frac{1}{\omega r}\right)\right],\label{I_modified}\\
    \tilde{K}_{l/\alpha}(\omega r)&=\left(\frac{\pi}{2\omega r}\right)^{1/2}e^{-\omega r}\left[1+O\left(\frac{1}{\omega r}\right)\right].\label{K_modified}    
\end{align}
Therefore, since the classical trajectories shown in Figs. \ref{space_geodesic} and \ref{space_polar} indicate the idea of bound states, a more appropriate physical solution is $\tilde{K}_{l/\alpha}$.

We are conscious that the model presented here has limitations concerning the cosmic singularity, particularly the radius of the inner core cylinder. However, it could be possible to use this as an advantage with an additional cost of a more developed metamaterial design. To see this, suppose that for the case in Fig. \ref{fig1} our new permittivities $\varepsilon_\parallel$, $\varepsilon_\perp$ are now functions of $r=\rho\sqrt{\epsilon_\perp}$ and given by
\begin{align}
	\varepsilon_\parallel(r)&=\epsilon_\parallel\left(1+\frac{\delta^2}{r^2}\right),\label{epsilon_radial}\\
    \varepsilon_\perp(r)&=\epsilon_\perp\left(1+\frac{\alpha^2\delta^2}{r^2+\delta^2}\right)\label{epsilon_circular},
\end{align}
where $\epsilon_\parallel$, $\epsilon_\perp$ are the previous values of the permittivities, $\alpha^2=|\epsilon_\parallel|/\epsilon_\perp$ and $\delta$ is the radius of the inner core, which is a small value. Substituting Eqs. (\ref{epsilon_radial}) and (\ref{epsilon_circular}) in the metric given by Eq. (\ref{effect_metric3}) leads to
\begin{equation}
	ds^2_\delta=dT^2-\left(1+\frac{\alpha^2\delta^2}{r^2+\delta^2}\right)dr^2+\alpha^2\left(r^2+\delta^2\right)d\phi^2.
    \label{rcm_effect_metric}
\end{equation}
Therefore, the wave equation becomes
\begin{multline}
	-\frac{r^2+\delta^2}{r^2+\delta^2(1+\alpha^2)}\frac{\partial^2\psi}{\partial r^2}-\frac{r[r^2+\delta^2(1+2\alpha^2)]}{[r^2+\delta^2(1+\alpha^2)]^2}\frac{\partial\psi}{\partial r}\\+\frac{1}{\alpha^2(r^2+\delta^2)}\frac{\partial^2\psi}{\partial\phi^2}-\omega^2\psi=0.
    \label{rcm_waveeq}
\end{multline}
Separating the variables $\psi(r,\phi)=f(r)g(\phi)$, one gets again Eq. (\ref{phi_sol}) for the angular solution. As for the radial part,
\begin{multline}
	\frac{(r^2+\delta^2)^2}{r^2+\delta^2(1+\alpha^2)}\frac{d^2f}{ dr^2}+\frac{r(r^2+\delta^2)[r^2+\delta^2(1+2\alpha^2)]}{[r^2+\delta^2(1+\alpha^2)]^2}\frac{df}{dr}\\+\left[\omega^2(r^2+\delta^2) +\frac{l^2}{\alpha^2}\right]f=0,
    \label{rcm_radial}
\end{multline}
which is more difficult to solve comparing with Eq. (\ref{ode_radial}). However, it can be done numerically as shown in Ref. \cite{poloneses2006}. Together with Eq. (\ref{phi_sol}) it represents the solution for a KG particle moving in a hyperboloid embedded in a 3D Minkowski space (with the usual change of variables $r\rightarrow t$, $\alpha\rightarrow \kappa$ as in the previous cases). This regularization was suggested in Ref. \cite{Turok_Singularity} and, in fact, used in Ref. \cite{poloneses2006} to remove the Cauchy problem at the singularity. The physical motivation is that, as the particle approaches the singularity, its own gravitational field modifies the spacetime, slightly deforming the double cone into a one sheeted hyperboloid (or similar surface). Thus, the extra space dimension does not contract to a point, but to some small value (represented above by $\delta$). As a result, the propagation is uniquely defined in the entire spacetime [the wave  functions are continuous at the origin (singularity)] in the same manner as for the case with $l=0$ described previously. We remark that the regularization was introduced in an \textit{ad hoc} manner and therefore  explicit gravitational backreaction calculations in the spirit of Refs. \cite{Poisson1,Poisson2004} are necessary to find the actual perturbed geometry of the Milne cone. With the simple regularization introduced above we see that,  through a more developed metamaterial design, it could be possible to circumvent or at least minimize the problem of the singularity in the optical analogy.

In this section we have seen that Klein-Gordon particles are annihilated upon reaching the singularity in the lower cone and created on the upper one, in agreement with the classical result of Sec. III.  The same happens to tachyons with the difference that they are in bound states, in accord with  the trajectories obtained in Sec. III.

As a last remark, we advise that there was an unfortunate error in Ref. \cite{fumeron2015optics}, where an  equation similar to Eq. (\ref{eq39}) was solved. The solution for that case also will be in the form of $\psi(r,\phi)=f(r)g(\phi)$, with $f(r)$ and $g(\phi)$ given by Eqs. (\ref{radial_sol}) and (\ref{phi_sol}), respectively.

\section{Conclusions}

The possibility of doing experiments in condensed matter systems that simulate cosmological cyclic/ekpyrotic scenarios was  proposed here with the particular focus on the Milne compactified universe $\mathcal{M}_C$, which is the simplest version of spacetime that can model the big crunch/big bang transition. This is the main goal of this work. Thus, we have shown that both Klein-Gordon particles and tachyons in $\mathcal{M}_C$  can be nicely represented by ordinary light propagating in specially engineered materials known as hyperbolic metamaterials. Within the realm of geometrical optics, we pointed out that the classical trajectories of those particles can be  perfectly matched to light ray paths in the metamaterial, while the quantum wave functions may be realized by wave optics. Furthermore, concerning the latter case, we pointed out that is possible to attenuate the problem of the singularity, designing a material whose  permittivity tensor has components which are functions of the radial variable $r$. We remark the exciting  possibility of experimentally checking, not only the trajectories, but also the lack of correlation between the wave function of  particles on both sides of the cosmological transition, through the analogue model presented here. Finally, further theoretical results  can     be found through the study of wave scattering, applying the method of partial waves \cite{erms/moraes2011} to the present model. This is presently under investigation and  will be the theme of a separate publication.

\begin{acknowledgments}
D.F. and F.M. are thankful for the financial support and warm hospitality  of the group at Universit\'e de Lorraine where this work was conceived and partly done. This work has been partially supported by CNPq, CAPES and FACEPE (Brazilian agencies).
\end{acknowledgments}

\bibliographystyle{ieeetr}
\bibliography{bib_milne.bib}

\end{document}